\documentclass[aps,pre,twocolumn,showpacs,amsmath,amssymb,floatfix,superscriptaddress]{revtex4}

\usepackage{graphicx}
\graphicspath{ {FIGURE/},{Figure/} } 

\begin{document}

\title{The LANER: optical networks as complex lasers}
\date{\today}

\author{Giovanni Giacomelli}
\affiliation{Consiglio Nazionale delle Ricerche, Istituto dei Sistemi Complessi, Via Madonna del Piano 10 I-50019 Sesto Fiorentino, Italy.} 
\author{Stefano Lepri}
\affiliation{Consiglio Nazionale delle Ricerche, Istituto dei Sistemi Complessi, Via Madonna del Piano 10 I-50019 Sesto Fiorentino, Italy.} 
\author{Cosimo Trono }
\affiliation{Consiglio Nazionale delle Ricerche, Istituto di Fisica Applicata "Nello Carrara", Via Madonna del Piano 10 I-50019 Sesto Fiorentino, Italy.}

\begin{abstract}
We discuss the main features of a new optical system capable of laser action: the active complex optical network, or lasing network (LANER). The system is experimentally realized with optical fibers linked each other with suitable optical couplers and with one or more coherent optical amplifying sections. The LANER displays a standard laser behavior: when the gain provided by the active sections is high enough to overcome the losses a coherent emission is produced, with a complicated intensity spectrum reflecting the structure of the network. A simple linear theoretical description is introduced and discussed, showing how the LANER can be considered as a generalization of the laser for a complicated cavity represented by the network itself. The system can be mapped to directed graphs and permits to disclosure the analogies with the problem of quantum chaos on graphs. In the case the links are all integer multiples of the same length, it is shown that the LANER framework corresponds to a lattice problem, with the equivalence of the Brillouin zone with the cavity Free Spectral Range.  Experimental realizations of different configurations are presented and examples of spectra are reported, in a phenomenological agreement with the numerical findings of the theory. 
\end{abstract}

\maketitle

\section{Introduction}

The concept of graph or network is central in complexity science. Countless examples of 
systems where non-regular connectivity of agents plays a central role in the dynamics
and on emergent properties are discussed in the scientific literature. 
In view of the many possible natural realizations in physics,
biology and even in social sciences, the study of nonlinear dynamical system
on graphs is \textit{per se} a relevant research topic \cite{Porter}.
Just to mention a few examples, this is relevant for the 
functioning of power grids and their failures 
\cite{Rohden2012}, the role of topology on synchronization \cite{belykh2005synchronization} or 
other nonlinear effects \cite{Gnutzmann2011,perakis2014small}. 
Dynamics of nonlinear fields on graphs, like for instance star-like structures, has also
achieved interest of theoreticians \cite{bellazzini2006,Noja2013}. However, 
most of the above topics could be quite difficult to be studied in real systems: even more, in controlled laboratory conditions.

Laser systems have been historically used as testbeds for many ideas from nonlinear dynamics and statistical or condensed-matter physics and could represent a suitable candidate for such an investigation. On the other hand, since the very first proposals, the great majority of lasers shared the same structure: a gain section in a simple linear or ring cavity, supporting regular sets of optical modes \cite{milonni2010laser}. The opposite case is represented by the random laser, where the propagation of rays in a disordered gain medium leads to light amplification \cite{Cao2003,Wiersma2008}. 

In fact, both such frameworks are rather inadequate to provide the richness and the complexity required for a characterization of specific statistical and/or dynamical issues in the network theory. In this work, we discuss in detail the recently introduced  \textit{lasing network} (LANER) \cite{lepri2017}, as a system where the above difficulties can be overcome. In brief, it consists of an active optical network, whose connectivity induces a form of \textit{topological disorder} and can display a genuine laser action; indeed, the LANER could also be considered as a \textit{discrete} random laser, with a controllable complexity \cite{lepri2017}. The system permits to scale the more standard laser geometries, embodied in the simpler configurations with a single gain, to strongly connected, multiple gain setups. Moreover, from an experimental point of view the apparatus has several practical
advantages: its flexibility that allow to explore different configurations by an easy re-arrangement of the components; the stability of the setup grants detailed statistical analysis as well. 

The plan of the paper is as follows. In Section \ref{sec:laner} we introduce the general idea and the main physical ingredients. A theoretical model leads to compute the linear modes via the network matrix of the LANER is presented in Section \ref{sec:theory}. Some specific examples are illustrated in Section \ref{sec:examples}, where we also report a calculation of the optical spectra based on rational approximations of the lengths (Section \ref{sec:lattice}). The final part of the paper (Sections \ref{sec:setup} and \ref{sec:exp}) is devoted to the experimental realizations and a summary of their phenomenology.
 
\section{The LANER concept}
\label{sec:laner}
\begin{figure}
\includegraphics[width=0.8\linewidth]{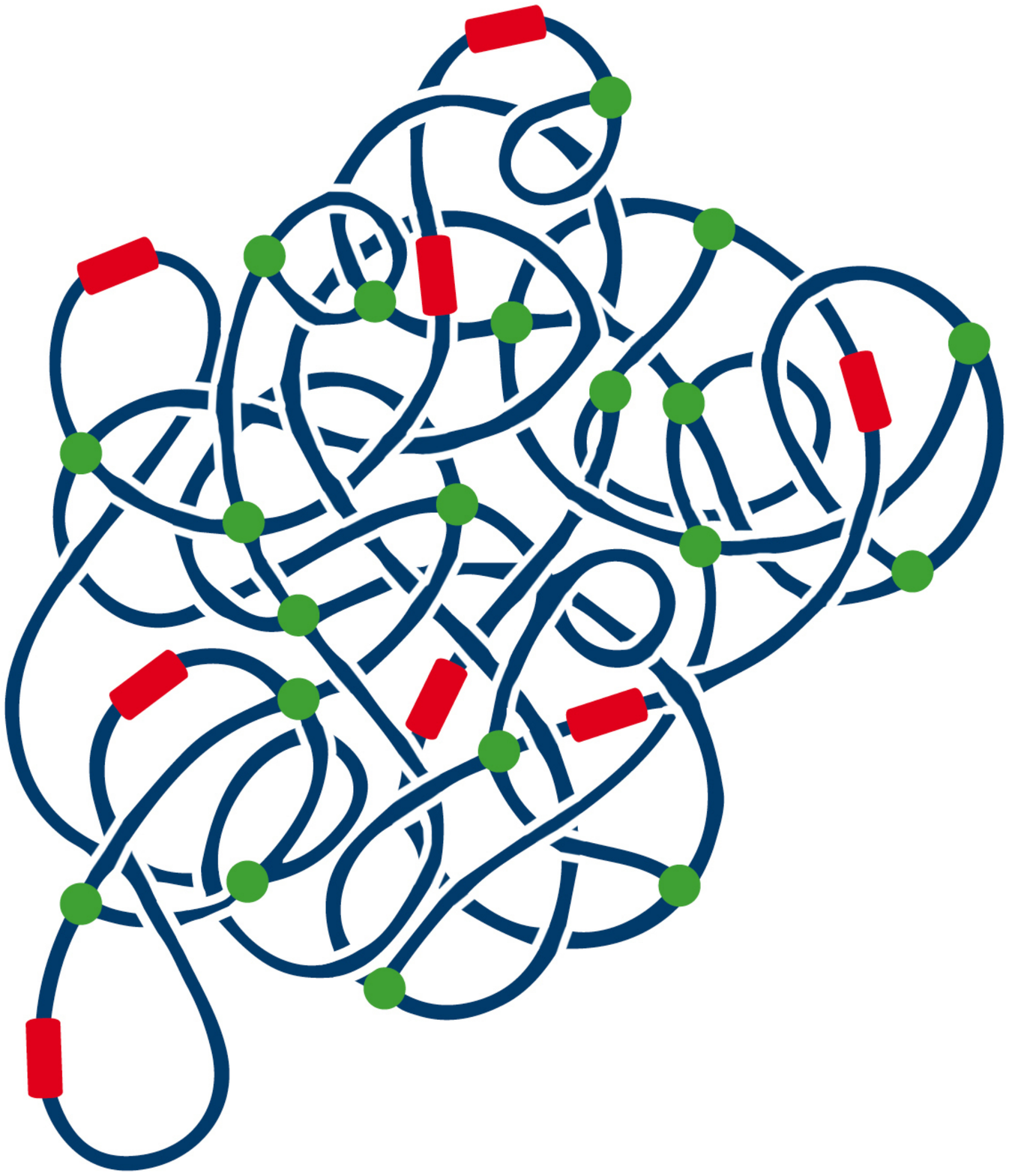}
\caption{The LANER concept. Optical fibers (blue), connected though standard components such as power splitters (green dots) form a closed network acting as a laser with a complex cavity. The coherent gains are provided by active regions (red thick segments) built e.g. with Er-doped fibers, semiconductor amplifiers, etc.}
\label{laner01}
\end{figure}

A sketch of the concept is depicted in Fig.\ref{laner01}. A network is built using single-mode optical fibers connected via standard optical components such as $m \times n$ power splitters, circulators, etc. Coherent gain sections, using the mechanism of stimulated emission are placed in one or more ({\it active}) links: laser-pumped, Erbium-doped fibers and/or semiconductor amplifiers or other systems can be used at this purpose. Alternative components can be employed to introduce/remove particular constraints in the structure of the network; for instance in the experiments described in this work we will only deal with {\it directed gains}, realized by inserting optical isolators in the active links. The observable quantities are the fields in the links, detected inserting $2 \times 2$ power splitters (possibly with a small coupling ratio):
the two propagation directions can be monitored at the same time if desired.
The split fields can be thus sent to high bandwidth/sensitivity detectors and the resulting intensity signal is proportional to the {\it beatings} of the optical modes. We notice how the detection arms, gathering optical power from the network represent unavoidable sources of losses.

\section{Theoretical description}
\label{sec:theory}
For the theoretical description, we denote by $L_j$ the lengths of the $N_l$
fiber segments; in a fully connected network $N_l = 2 N_s$ for the splitters we consider in the present work. The main basic quantity we will deal with is the optical spectrum of stationary modes of the network. Its calculation is based on a standard linear propagation approach and will be discussed in the following.

\subsection{The observables: the field in the links}

\begin{figure}
\includegraphics[width=1.0\linewidth]{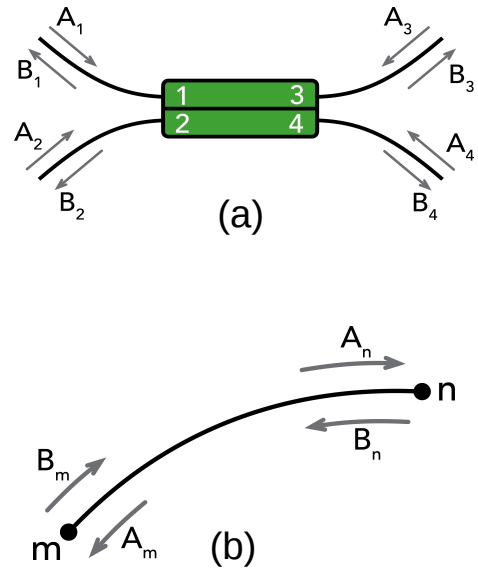}
\caption{The observables: the propagating fields at the ports of the splitters (a) and at the beginning of the links (b).}
\label{observ02}
\end{figure}

We consider the electric field propagating in the link $j$
\begin{equation}
{\bf E}_j(t,{\bf r}_j) = {\bf E}_j^{(+)}(t,{\bf r}_j) + {\bf E}_j^{(-)}(t,{\bf r}_j)~,
\end{equation}
in terms of the forward/backward propagating fields
\begin{equation}
{\bf E}_j^{(\pm)}(t,{\bf r}_j) = {\bf e}_y \Phi(\rho,\theta)f_j^{(\pm)}(t,z_j)~.
\end{equation}
Here, ${\bf r}_j = (\rho,\theta,z_j)$ is the (local) cylindrical-coordinate vector, ${\bf e}_y$ the unit vector indicating the linear (transverse) polarization directions and $\Phi$ is the transverse mode profile (in single mode fibers, only the fundamental mode $LP_{01}$ function of the radial coordinate $\rho$ is usually supported; in our case, it is assumed to be the same in all the network). The functions $f^{(\pm)}(t,z_j)$ describe the amplitude of the field propagating in the two directions at the point $z_j$ of the $j$-link coordinate system. 

In the following, we will consider a single polarization direction in the whole network. While this can be experimentally realized by using polarization-maintaining fibers and components, the effect of possible different polarization directions in the network will be taken in account by a re-definition of the gains (see below). This approach is made possible by the fact that such effects are quenched, i.e. fixed for a given experimental configuration.  

The description of the LANER is then carried out using the above propagating fields in each link of the network. In particular, for a given link, we call $A$ (resp. $B$) the values of the field at the extremities exiting (resp. entering) the link as shown in Fig.\ref{observ02}b.

We remark how a general treatment should consider a specific description of the active media, with a proper analysis of the involved population(s) dynamics and keeping in account the propagation delays in there as well. Such an approach is currently under development and will be published elsewhere.

\subsection{Connecting the links: the scattering matrix}

We consider first a single 2$\times$2 loss-less coupler depicted in Fig.\ref{observ02}a.
We define, at each of the four ports, the column vectors of input and output amplitudes 
$A = (A_1,A_2,A_3,A_4)^T$ and  $B = (B_1,B_2,B_3,B_4)^T$ (this requires fixing 
a port numbering). They are related via the scattering matrix $B = S^{(1)}A$.
The splitting of the field follows an unitary transformation (energy conservation through the splitter). Under a prescribed convention, we can parametrize it through 
the transmission and reflection coefficients $T,R$ such that $|T|=\sqrt{\alpha}$ and  
$|R|=\sqrt{1-\alpha}$ where $\alpha$ is also termed 
as the \textit{splitting factor}. 
Moreover, the splitters are reflectionless and this 
impose some further  constraint on the form $S^{(1)}$. Altogether
\begin{equation}
S^{(1)} = \left( \begin{array}{cccc}
0 & 0 & \sqrt{1-\alpha} & \sqrt{\alpha} \\
0 & 0 & -\sqrt{\alpha} & \sqrt{1-\alpha} \\
\sqrt{1-\alpha} & -\sqrt{\alpha} & 0 & 0 \\
\sqrt{\alpha} & \sqrt{1-\alpha} & 0 & 0 \end{array} \right) 
\label{S}
\end{equation}
The blocks connecting $\{A_1,A_2\}$ with $\{B_3,B_4\}$ and $\{A_3,A_4\}$ with $\{B_1,B_2\}$ (upper right and bottom left respectively) are unitary matrix as noted before. 
As a consequence, the matrix $S^{(1)}$ is unitary as well.

In the case of $N_s$ splitters that, for simplicity, we assume to be all equal as above, 
we can extend the definitions introducing  
the $4N_s$-dimensional column vectors of input and output amplitudes 
$A$ and $B$ whose components are taken ordered in such a way that 
\begin{equation}
B = S A
\label{Smat}
\end{equation}
and the complete ($4 N_s\times4 N_s$) scattering matrix is  block-diagonal
\[ S  =
 \begin{pmatrix}
  S^{(1)} & 0 & \cdots & 0 \\
  0 & S^{(1)} & \cdots & 0 \\
  \vdots  & \vdots  & \ddots & \vdots  \\
  0 & 0 & \cdots & S^{(1)}
 \end{pmatrix} ~. \]
The unitary (orthogonal) matrix $S$ describes the transfer properties of the 
splitters (or, in general, of the optical components at the nodes of the network) 
and is thus independent of the gains 
and the topology of the network. It is readily generalized to the case
of a set of different splitters each having different transmission coefficients and/or 
port numbers. 

It is worth noting that splitters may have rather different design translating into corresponding properties of $S$, such as (see e.g. \cite{Pozar}) 
\begin{enumerate}
\item[$S_1$:] {\it lossless} (S is unitary);
\item[$S_2$:] {\it reciprocal} (S is symmetric);
\item[$S_3$:] {\it matched}  (diagonal elements of S are zero).
\end{enumerate}

In the present work we will focus on the case of all equal $2\times2$, $50\%$ power optical splitters, such that $\alpha = {1\over2}$ which satisfies $S_1$, $S_2$ and $S_3$.

\subsection{Evolution in the links: the propagation matrix}
\label{subsec:prop}

In a linear description of the network, we assume that the field in the link $j$ is multiplied by a factor ({\it gain}) $G_j = g_j e^{-(\mu +i k) L_j} = g_j e^{-s L_j}$, where $g_j>1$  (resp $<1$)  represents real and positive link gain (resp. losses), $s$ the complex wave-vector and $L_j$ the (oriented) link length. According to the definition of the field variables, and introducing  the $(2N_l\times 2N_l)$ {\it propagation} matrix $P$, we write
\begin{equation}
A = P B~,
\label{Pmat}
\end{equation}
where in a fully connected network $N_l = 2 N_s$; 

In a specific realization of the LANER, additional (fixed) link/coupler phase delays should be considered depending on the type of components. Such delays could be included in a more general form of the gains as 
\begin{equation}
G_j^{(\pm)} \to G_j^{(\pm)} e^{i \phi_j^{(\pm)}}~,
\label{cgains}
\end{equation}

where the ($\pm$) refers to the (chosen) propagation direction along the link: notably, $\phi_j^{(+)} = -\phi_j^{(-)}$.
Associating the global phase delays to the elements of $P$ allows to write the matrix $S$ in its simplest form, i.e. possible phase delays due to the couplers are now included in the propagation and not in the scattering matrix.

The matrix $P$ contains both the topology (how the splitters connect the links) and metric (gain and links length) information of the network.
As shown in Fig.\ref{observ02}, variables $\{A,B\}$ in the same link are related according to (we have not made explicit the link index for simplicity)
\begin{eqnarray}
A_n &=& g^{(+)} e^{-(\mu +ik) L} B_m= G^{(+)}~B_m\\
A_m &=& g^{(-)} e^{-(\mu -ik) L} B_n= G^{(-)}~B_n~.
\end{eqnarray}

When $g^{(+)} = g^{(-)} = g$,
\begin{equation}
P_{nm} = g e^{-s L} = G~,~P_{mn} = G^*~;
\label{ggains}
\end{equation}
we will most often consider the case $\phi = 0$ for each link, since in our experiments phase delays are not present either in the fibers or in the power splitters. 

Directed active gains, that we are using in the experiment reported here in fact represent a remarkable exception of the above situation. In this case, the blocked direction in the symmetric elements of $P$ will be set to zero: notably, this is a strong source of losses, as the field incoming in the blocked direction is dissipated in the optical isolator.

The matrix $P$ has the following properties:
\begin{enumerate}
\item since every port of each splitter is connected to one and only one (different) port,  
$P$ has only one nonvanishing entry in every row and in every column;

\item since the output of a port cannot be connected to itself, the diagonal elements 
of $P$ are zero; 

\item it is invertible since 
\begin{eqnarray*}
&&\det P =  |G_1|^2\times..\times|G_{N_l}|^2 = \\ 
&&= (g_1^{(+)} g_1^{(-)})\times...\times (g_{N_l}^{(+)} g_{N_l}^{(-)})\times \exp (-2\mu L) \ne 0
\end{eqnarray*}
where we have introduced the total length $L=L_1+...+L_{N_l}$;

\end{enumerate}

In addition, in the particular case of symmetric gains $g_j^{(+)}=\big( g_j^{(-)} \big)^*$ (i.e. $|g_j^{(+)}|=|g_j^{(-)}|$ since $\phi_j^{(+)} = -\phi_j^{(-)}$ as said) for every $j$, $P$ is {\it hermitian} (i.e. $P = P^\dag$).

\subsection{The network matrix}

The field stationary solution $A$ in the network can be obtained  by
writing
\begin{equation}
A = P B = P S A~.
\label{stat}
\end{equation}
The equation (\ref{stat}) is satisfied if $A$ is an eigenvector of the {\it network} matrix 
\begin{equation}
N=PS~,
\label{netm}
\end{equation}
with a corresponding eigenvalue $\lambda = 1$, leading to the characteristic equation  \footnote{Eq.(\ref{stat}) can be written in terms of $B$ as well, using the matrix $M = SP$. In particular, the eigenvalues of $N$ and $M$ are the same.} 
\begin{equation}
\det\Big(N-I\Big)=0~,
\label{char}
\end{equation} 

where $I$ is the identity matrix.
Eq.(\ref{char}) allows for the complete determination both of the optical modes supported by the cavity (matching conditions for the field) and of the threshold condition (balance between the gains and the losses in the network) in order to have them lasing. Analytical solutions of (\ref{char}) are often very difficult to be found except for some simple cases that we will examine  below. The roots $s$ in the complex plane 
can be found numerically by standard routines. 

We remark how, since the direction in every link is chosen arbitrarily, Eq.\ref{char} must admit the same set of solutions for all the possible permutations $G_j^{(+)} \to G_j^{(-)}$ ($G_j \to G_j^*$ in the case of symmetric scalar gains).  

Depending on the configuration (topology of the connections) we expect that the corresponding matrix $N$ might exhibit particular structures allowing the determination of the properties of the eigenvalues. The above relation represents the mayor challenge of the linear, theoretical description of a LANER.

\subsection{Lossless network (empty cavity limit): the hamiltonian case}

Let us now discuss two particular cases, starting with the situation of purely passive network
(no gain nor lossess), i.e. $g_j =1$. This corresponds to the standard 
empty cavity limit, and can be described mathematically by setting the elements 
of the propagation matrix to pure dephasing terms only: $G_j \rightarrow e^{iKL_j}$
(with $K$ real). 
Here the propagation matrix is unitary 
$P^\dagger=P$; this reflects time-reversal invariance of the dynamics (which is obviously
broken in presence of gain and losses). Thus the network matrix is unitary as well,
$N^\dagger N=N N^\dagger = I$.
In this case, all the modes are marginally stable and the dynamics would be akin
to the one of a conservative system whereby all the possible modes can lase 
and the equation provides a way for calculating 
all of them regardless of their effective action in the dynamics.

We remark also the close analogies of this situation with \textit{quantum graphs} which 
have been thoroughly investigated in the realm of quantum chaos 
\cite{kuchment2004quantum,kottos1999periodic,kottos2003quantum}. In this context $N$ is termed the vertex scattering matrix. The equation for the poles, 
Eq.\ref{char}, is formally equal to the one to determine the quantum spectrum of a particle moving freely along the bonds and scattered at the graph vertices. As an interesting 
extension that may be of some relevance also in our case, we mention that a form of dissipation 
has been also studied by considering open graphs \cite{barra2001transport,kottos2003quantum}.
 
An experimental realization of quantum graphs is provided by microwave networks which
have been investigated in recent years \cite{hul2004experimental}. However, the LANER 
is not only different for being an optical system but also for having as a novel
element the optical gain (possibly in multiple links) which allow to achieve the lasing 
action and investigate entirely new effects.

\subsection{Mapping the LANER on graphs}

Another interesting limit is the one in which the propagation approaches a "classical" limit
corresponding to ray dynamics \cite{barra2001classical}. This point of view has been successfully 
employed to justify theoretically the L\'evy distribution of emission fluctuations, akin
to the one observed for random lasers \cite{Ignesti2013,Uppu2014,lima2017}. 
To this aim, a Monte-Carlo type of model has been simulated in Ref.\cite{lepri2017}. 
It consists of two steps: (i) a free propagation of an ensemble of
"rays" on the physical graph, each having an intensity that grows or 
decreases depending on the local value of the gains $g_j$ and (ii) random transitions 
at each splitter with probabilities assigned according to the values of the splitting factors (see Ref.\cite{lepri2017} for details).

In this situation, it should be also useful to introduce a simplified description of the dynamics as 
a (random) motion on an equivalent graph that can be introduced as follows. 
Each vertex of the graph represents a \textit{component of the field} in a link, and a directed 
bond the linear dependence of the target on the source field. 
We define such a graph through its adjacency matrix obtained by setting all 
the nonzero elements of the matrix $N$ to 1. 

\begin{figure}
\includegraphics[width=1.\linewidth]{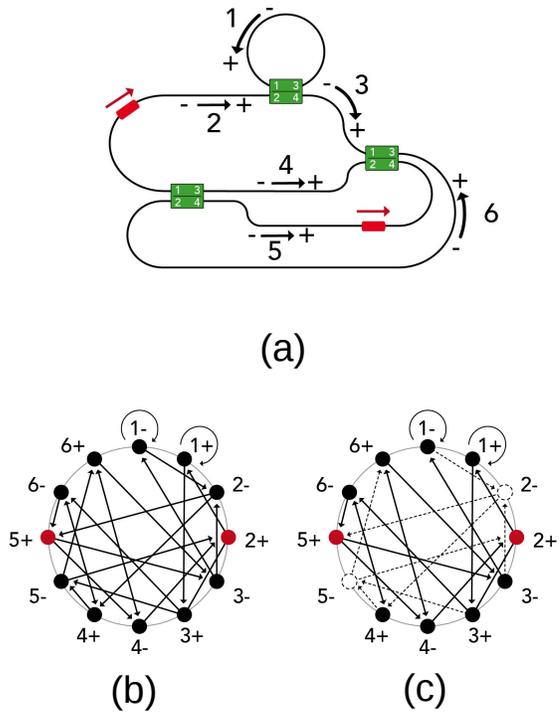}
\caption{A possible LANER configuration and its equivalent directed graph as described in the text. (a) The physical network: 
the thick, red segments indicate the oriented gain sections and the black, numbered arrows the propagating fields along the fiber links. (b) The graphs represent the fields as vertices and the connections induced by the splitters as bonds: a targeted mode is linearly dependent on the sourced one. The red 
and black dots indicate that the links are respectively active or passive. 
(c) change in the connectivity induced by the insertion of isolators as 
depicted in (a): the graph is pruned as the opaque parts are removed.}
\label{graph03}
\end{figure}

To illustrate the concept, we refer to the specific example with $N_l=2N_s=6$ sketched 
in Fig.\ref{graph03}(a). The numbers $1_\pm,2_\pm \ldots, 6_\pm$ denote the 12 field
components propagating on each fiber and label the nodes of the equivalent graph, Fig.\ref{graph03}(b). This example leads to a regular directed graph as the number of incoming and outcoming link is equal to 2 for each vertex. 
An red vertex indicates that the related mode link is active, otherwise passive if black. 
To illustrate how the connectivity can be changed, we show in  Fig.\ref{graph03}(c) how 
the inclusion of isolators in the active links leads to the removal of the blocked fields and therefore the corresponding vertices are removed from the equivalent graph, which is therefore {\it pruned}. It is worth noting that the system can lase only if at least one active vertex is eventually present in the pruned graph. 

At the simplest level of description, the ray dynamics in the LANER can be visualized as 
chaotic motion yielding a  
Markovian random walk \cite{Burioni2005} on such graphs while amplification and 
damping can be introduced as in the model mentioned here above. 

\section{LANER examples}
\label{sec:examples}
In the following Sections we consider the simpler configurations, realized with few splitters in the $50:50$ splitting ratio case, i.e. $\alpha=1/2$ in Eq.(\ref{S}) and symmetric scalar gains. We notice how the zero splitter LANER is the standard ring laser.

\begin{figure*}
\includegraphics[width=.9\linewidth]{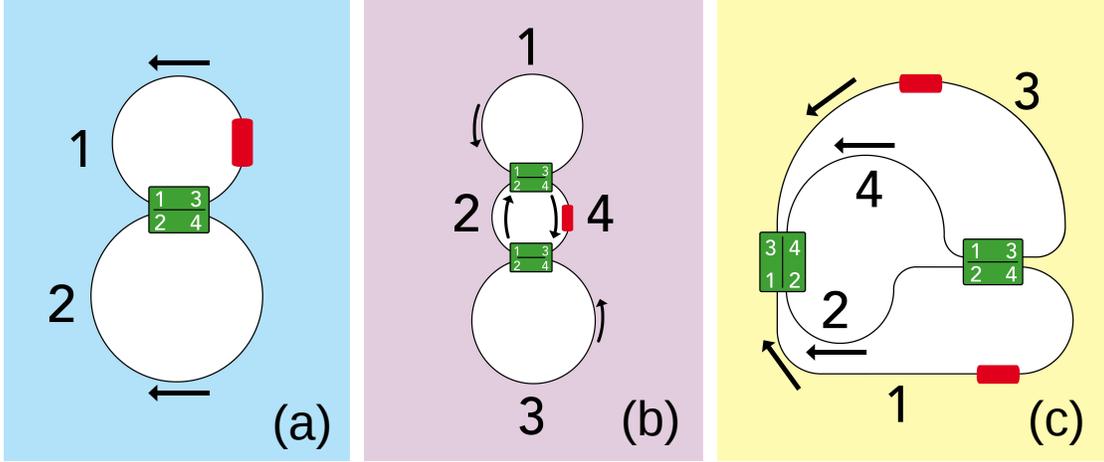}
\caption{LANER examples: (a) 1PARA, (b) 2PARA and (c) 2PERP. The arrows indicate the (arbitrarily chosen) reference propagation directions. Red blocks depict possible position choices for the active gains.}
\label{configs04}
\end{figure*}

Starting with the case of a single splitter, we consider the configuration 1PARA 
represented in Fig.\ref{configs04}a as the simplest, non-trivial LANER.
 The propagation matrix is

\begin{equation}
P = \left( \begin{array}{cccc}
0 & 0 & G_1 & 0 \\
0 & 0 & 0 & G_2 \\
G_1^* & 0 & 0 & 0 \\
0 & G_2^* & 0 & 0 \end{array} \right)~.
\end{equation}
The gains are oriented (here and in the following examples) according to the arrows depicted in the figure, i.e. the propagation gain along the arrow is $G$ and $G^*$ in the opposite direction.

The network matrix is
\begin{equation}
N = PS = \frac{1}{\sqrt 2} \left( \begin{array}{cccc}
G_1 & -G_1 & 0 & 0 \\
G_2 & G_2 & 0 & 0 \\
0 & 0 & G_1^* & G_1^* \\
0 & 0 & -G_2^* & G_2^* \end{array} \right)~. 
\end{equation}
The characteristic equation (\ref{char}) in this case factorizes as 
\begin{equation}
 \big(1+G_1^*G_2^* - \frac{1}{\sqrt{2}}(G_1^*+G_2^*)\big)\big(1+G_1G_2 - \frac{1}{\sqrt{2}}(G_1+G_2)\big)=0~.
\end{equation}

In the case of two splitters, more configurations can be realized; we first study the 2PARA, depicted in Fig.\ref{configs04}b. 
The associated propagation matrix is 
\begin{equation}P=\left(\begin{array}{cccccccc}
0 & 0 & G_{1} & 0 & 0 & 0 & 0 & 0\\
0 & 0 & 0 & 0 & G_{2} & 0 & 0 & 0\\
G_{1}^* & 0 & 0 & 0 & 0 & 0 & 0 & 0\\
0 & 0 & 0 & 0 & 0 & 0 & G_{3}^* & 0\\
0 & G_{2}^* & 0 & 0 & 0 & 0 & 0 & 0\\
0 & 0 & 0 & 0 & 0 & 0 & 0 & G_{4}^*\\
0 & 0 & 0 & G_{3} & 0 & 0 & 0 & 0\\
0 & 0 & 0 & 0 & 0 & G_{4} & 0 & 0
\end{array}\right)\end{equation}
and network matrix
\begin{equation*}N=
 \frac{1}{\sqrt{2}}\left(\begin{array}{cccccccc} 
 G_{1} & -  G_{1} & 0 & 0 & 0 & 0 & 0 & 0\\
 0 & 0 & 0 & 0 & 0 & 0 &  G_{2} &  G_{2}\\
 0 & 0 &  G_{1}^* &  G_{1}^* & 0 & 0 & 0 & 0\\
 0 & 0 & 0 & 0 &  G_{3}^* & -  G_{3}^* & 0 & 0\\
 0 & 0 & -  G_{2}^* &  G_{2}^* & 0 & 0 & 0 & 0\\
 0 & 0 & 0 & 0 &  G_{4}^* &  G_{4}^* & 0 & 0\\ 
 G_{3} &  G_{3} & 0 & 0 & 0 & 0 & 0 & 0\\
 0 & 0 & 0 & 0 & 0 & 0 & -  G_{4} &  G_{4}
\end{array}\right)~.
\end{equation*}

Another, non-trivial configuration is the 2PERP presented in Fig.\ref{configs04}c, described by the propagation matrix 
\begin{equation}
P=
\left(\begin{array}{cccccccc}
0 & 0 & 0 & 0 & 0 & 0 & 0 & G_{1}\\
0 & 0 & 0 & 0 & 0 & G_{2} & 0 & 0\\
0 & 0 & 0 & 0 & 0 & 0 & G_{3} & 0\\
0 & 0 & 0 & 0 & G_{4} & 0 & 0 & 0\\
0 & 0 & 0 & G_{4}^* & 0 & 0 & 0 & 0\\
0 & G_{2}^* & 0 & 0 & 0 & 0 & 0 & 0\\
0 & 0 & G_{3}^* & 0 & 0 & 0 & 0 & 0\\
G_{1}^* & 0 & 0 & 0 & 0 & 0 & 0 & 0
\end{array}
\right)
\end{equation}

with the network matrix
 \begin{equation*}
 N=
 \frac{1}{\sqrt{2}}\left(\begin{array}{cccccccc}
 0 & 0 & 0 & 0 &  G_{1} &  G_{1} & 0 & 0\\
 0 & 0 & 0 & 0 & 0 & 0 & -  G_{2} &  G_{2}\\
 0 & 0 & 0 & 0 &  G_{3} & -  G_{3} & 0 & 0\\
 0 & 0 & 0 & 0 & 0 & 0 &  G_{4} &  G_{4}\\ 
 G_{4}^* &  G_{4}^* & 0 & 0 & 0 & 0 & 0 & 0\\
 0 & 0 & -  G_{2}^* &  G_{2}^* & 0 & 0 & 0 & 0\\ 
 G_{3}^* & -  G_{3}^* & 0 & 0 & 0 & 0 & 0 & 0\\
 0 & 0 &  G_{1}^* &  G_{1}^* & 0 & 0 & 0 & 0
 \end{array}\right)~.
 \end{equation*}

We mention that, for the simple configurations such as the 1PARA and 2PARA, the equivalent graphs are {\it disjoint} \cite{lepri2017} indicating that two sets of independent propagations are possible in such LANERs. Such a feature in terms of ray dynamics can be interpreted as the independence of the two correspondent Euler paths \cite{chartrand1985introductory} in the graph.

\begin{figure*}
\includegraphics[width=0.3\linewidth]{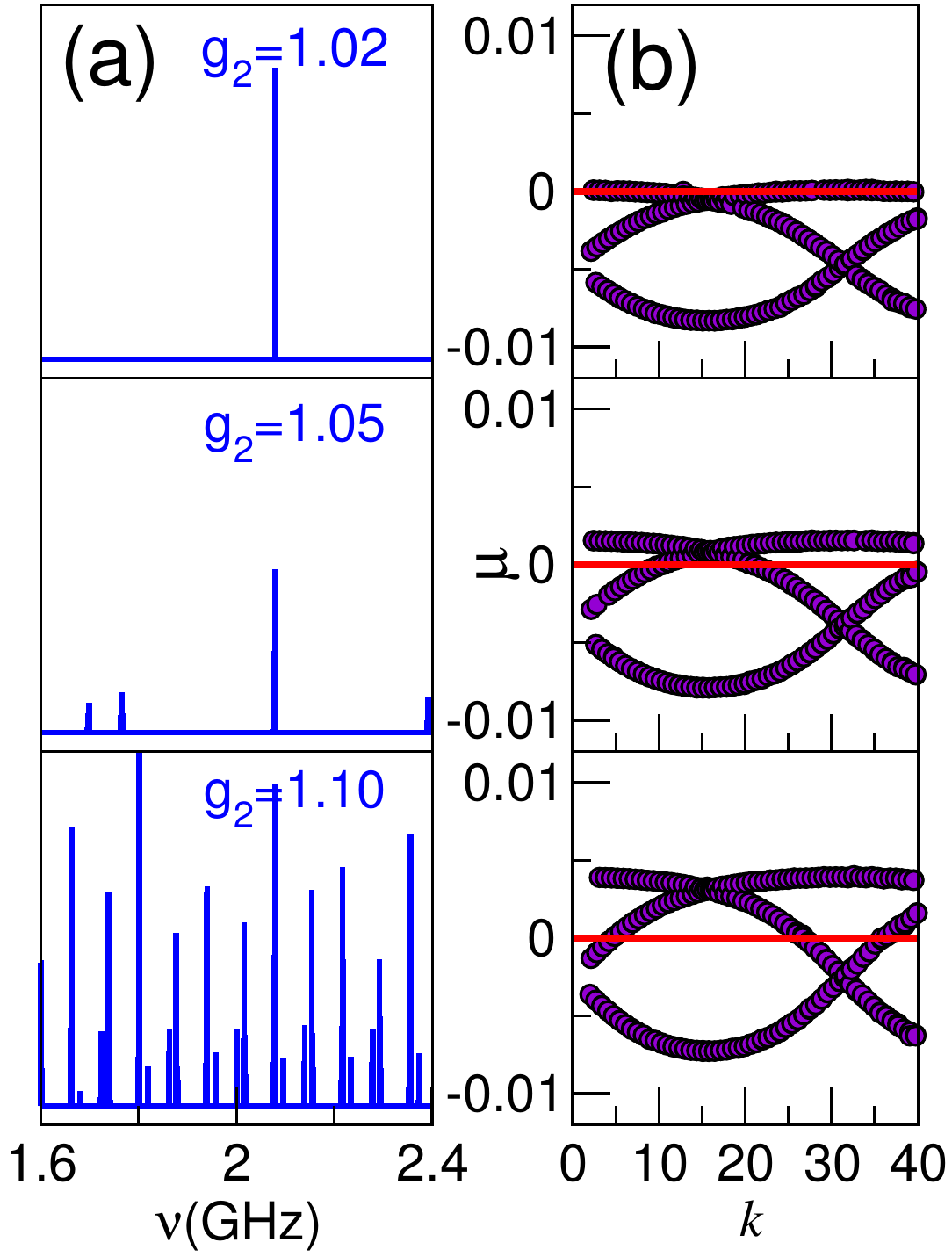}
\includegraphics[width=0.3\linewidth]{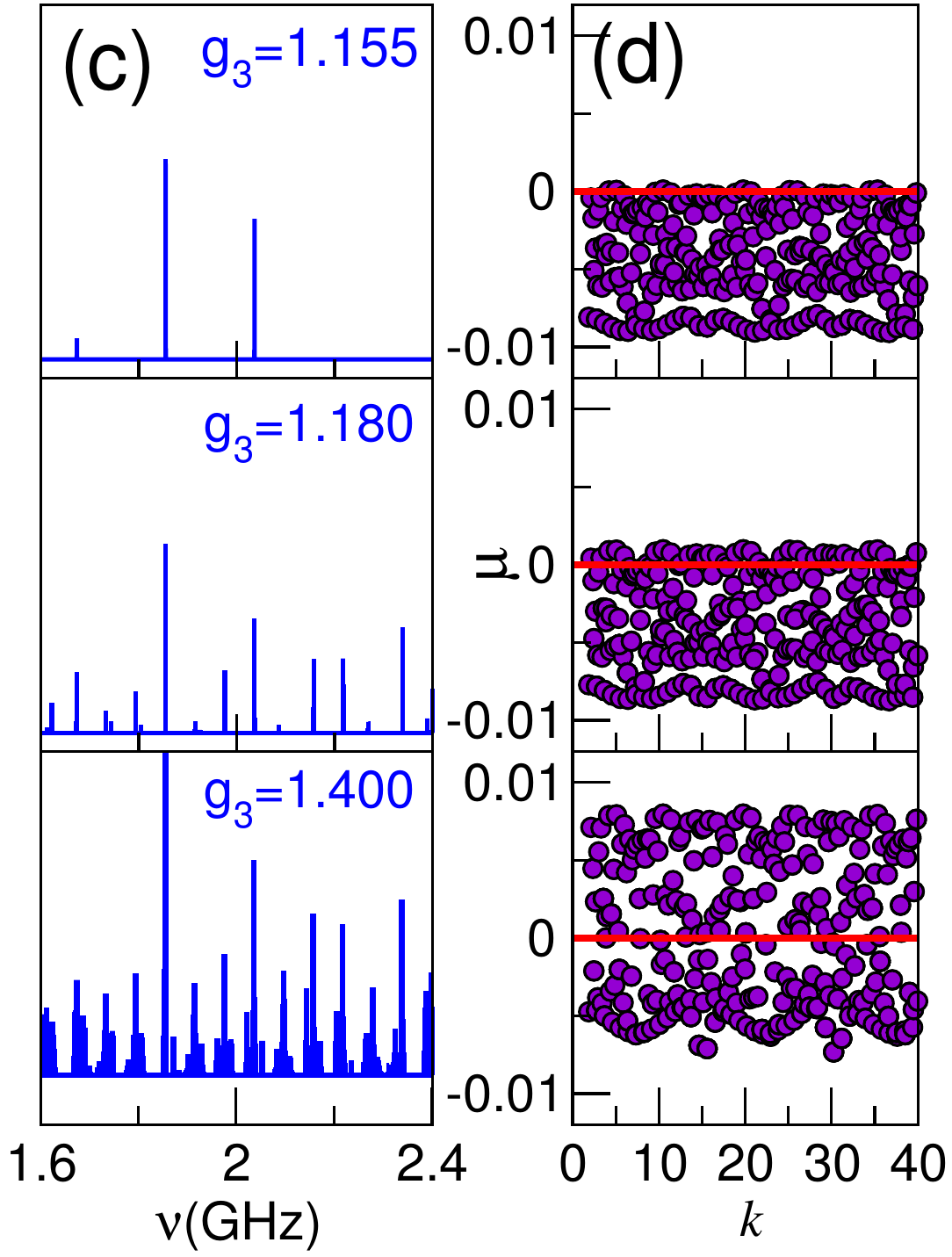}
\includegraphics[width=0.3\linewidth]{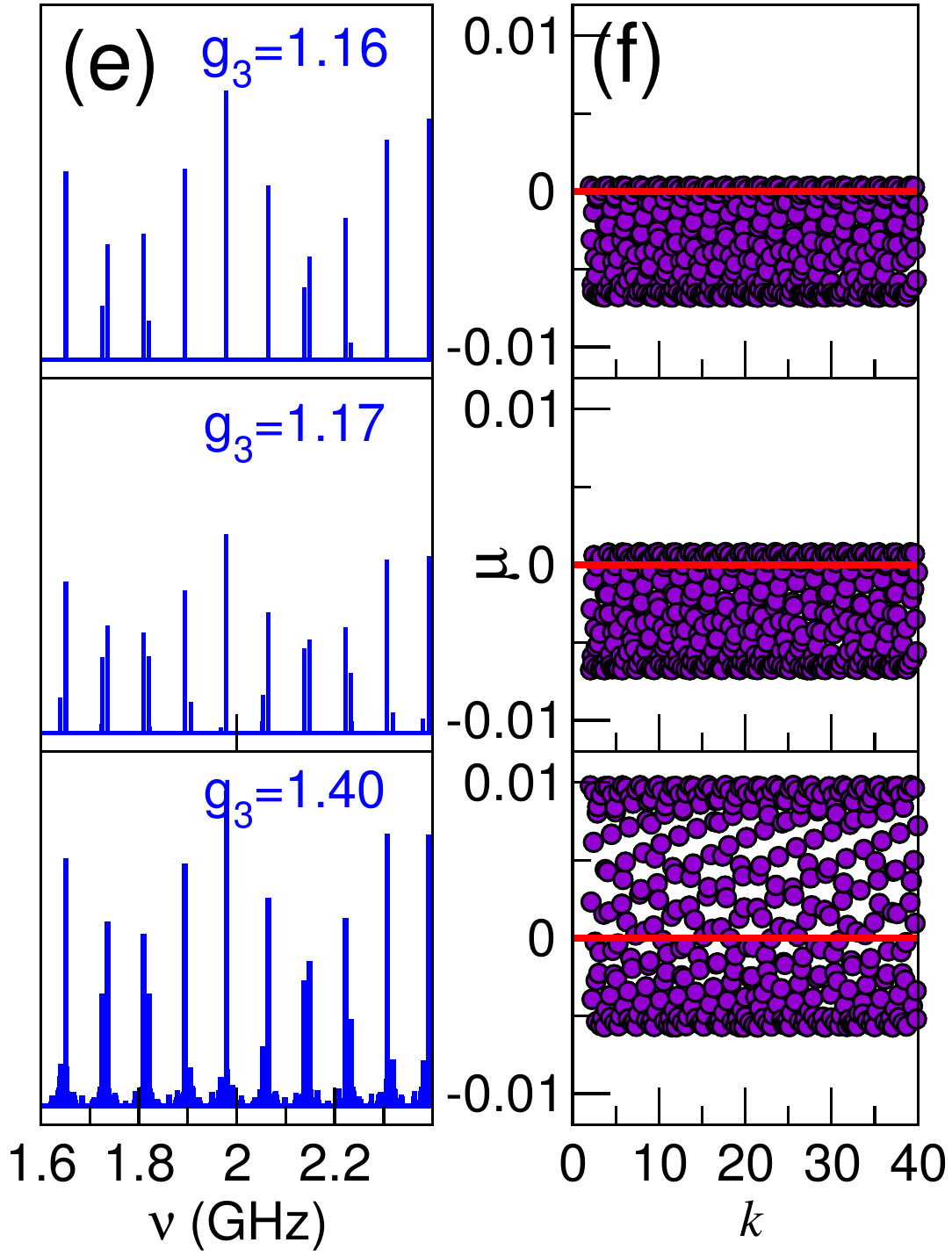}
\caption{(Numerics) Threshold behavior of various LANER systems upon increasing
gain on one link.
Results for 1PARA (left), 2PARA (center) 2PERP (right).
Panels (a,c,e) are the beating spectra (in arbirtary units) computed as described in the text; 
(b,d,f) poles in the complex plane. The chosen lengths are $L_1=9.16m,L_2=18.12m$ (1PARA) and 
$L_1=9.16m,L_2=18.12m,L_3=5.24m,L_4=10.0m$ (2PARA,2PERP), the loss factors are $g_l=0.9$
on all the links except the one specified in the figures.
The spectra are computed considering roughly $10^3-10^4$ poles in each case.}
\label{fig:soglia05}
\end{figure*}

Let us illustrate the spectral and emission features for the three topologies.
We first show the onset of the laser action by computing the poles $s_n=\mu_n +ik_n$ numerically 
from Eq.~(\ref{char}) and plotting them in the complex plane  
for increasing gain values (see Fig.\ref{fig:soglia05} b,d,f). In all cases,  
above a critical value, a set of resonances cross the real axis and the associated modes grow with rate 
$\mu_n>0$ and start to lase. Since we assumed an infinite gain bandwidth, the dynamics above threshold will be very high-dimensional. Also, we checked that the integrated density of states, namely the number of modes below $k$ grows on average linearly, 
as expected from the Weyl law for one-dimensional structure
\citep{kottos1999periodic}.

Since what is experimentally accessible is the beating spectra,    
for a given set of $M$ values of the wavenumbers $k_n$ we 
compute all the differences $k_n-k_m$ belonging to a given interval $(0,B)$, with $B$
being an assigned bandwidth. The histogram of the beatings obtained from the computation of the poles $s_n$: all the possible differences in frequency are evaluated for the lasing modes (those with $\mu_n>0$)  and their distribution is plotted (see Fig.\ref{fig:soglia05}~a,c,e). The result gives a qualitative estimation of the beating spectrum, showing a complex hierarchical structure that can be directly compared to the experiment (see later Sections).

\section{The network as a lattice}
\label{sec:lattice}

An important case to consider is when the links are all multiples of the same length $L_0$, i.e. the network is a {\it lattice} and $L_0$ is the lattice periodicity:
\begin{equation}
L_j = n_j L_0~,
\label{lattice}
\end{equation}
with $\{n_j\}$ being a set of positive integers.
Such a circumstance is intentionally avoided in the quantum graphs literature as it may lead to dynamical quasi-periodicity that hinders the observation of universal spectral features \citep{kottos1999periodic} but it is very instructive to gain insight in the spectral structure as we show in the following. 

According to Eq.~(\ref{lattice}), the gain terms turn to be invariant under the transformation $s \rightarrow s + iK_0$, where $K_0 = 2\pi/L_0$. The spectrum is therefore periodic along the frequency axis and the period $K_0$ is the Free Spectral Range (FSR, see e.g. \cite{milonni2010laser}) of the network cavity: the above result amounts to identify the FSR with the first Brillouin zone in solid state physics (see e.g. \cite{ashcroft}). In particular, one can consider the first zone as $[-\pi/L_0,\pi/L_0]$ and, since if $s$ is a solution $s^*$ is a solution as well it follows that the independent solutions are only those in $[0,\pi/L_0]$. 

Another remarkable feature is that in this case the poles can be evaluated as zeros of 
a polynomial and can be thus enumerated exactly. To see this, let's introduce the complex variable 
$e^{-s L_0} = z$ and define 
\begin{equation}
Q(z,z^*) = \det\Big(N-I\Big)~.
\end{equation} 
In the case of symmetric scalar gains, by inspection of $Q$ one can find that all its terms are of the form 
\begin{equation}
a_j z^k (z^*)^l +a_j^* z^l (z^*)^m ~,
\end{equation}
where the coefficients $a$ are products of the $g$'s. Indeed, since in our case $P$ is hermitian 
and $S$ is real and symmetric, it holds that
\begin{eqnarray}
Q^* &=& \det\Big(P^*S^*-I\Big)\\ \nonumber
    && =\det\Big( (SP)^T-I\Big)=\det\Big( SP-I\Big) = Q~.
\end{eqnarray}

$Q$ is thus a {\it real} polynomial, with a degree $D = \sum_l n_l$ both in $z$ and $z^*$ and admits $D$ solutions in the Brillouin (half) zone. The full solution for the poles $s$ are obtained by periodic extension taking all branches of the logarithm function with the appropriate periodicity.

We notice that in general the characteristic equation
\begin{equation}
\det\Big(N(G,G^*)-I\Big) = C(G,G^*)= 0~,
\end{equation}

must admit the same set of solutions not only by replacing $G_j \leftrightarrow G_j^*$ for all $j$ (to invert all the propagation direction in the links), but also in the case of the more general symmetry 
\begin{equation}
G_j \leftrightarrow G_j^*,~j \in H~, 
\end{equation}

where $H$ is an arbitrary subset of the links index set. Such a property is direct consequence of the possibility to choose arbitrarily the propagation direction in every link. Besides, it is always possible to write the propagation matrix as
\begin{equation}
P(G,G^*) = T_1(G,G^*) +T_2(G,G^*),
\end{equation}

where $T_1$ (resp. $T_2$) is a lower (upper) triangular matrix and, with a proper choice of the orientation in the links we can order the gains in such a way that e.g. all and only the $G$'s (resp. $G^*$'s) are $T_1$ ($T_2$) elements. Finally, using $P=P^\dag$ it follows that
\begin{equation}
P(G,G^*) = T(G) +T(G)^\dag~,
\end{equation}

where $T = T_1$. The characteristic equation thus can be solved by considering
\begin{equation}
C(G,G)= 0~,
\end{equation}
together with the symmetry $G \leftrightarrow G^*$ (if $G$ is a solution $G^*$ is a solution as well).

In the following, we discuss as an example the 1PARA spectrum in the lattice configuration. We first set 
\begin{equation}
L_1 = n_1 L_0, ~L_2 = n_2 L_0 ~;
\label{rational2}
\end{equation}
to find the poles of the characteristic equation is sufficient to solve
\begin{equation}
1+g_1g_2 z^{n_1+n_2} - \frac{1}{\sqrt{2}}(g_1 z^{n_1} +g_2 z^{n_2})=0~,
\label{poly}
\end{equation}
where $z = e^{-sL_0}$, given the spectrum symmetry discussed above. 

For illustration, let us first consider the case $n_2 = N n_1$, 
where now $z = e^{-s n_1 L_0}$ yielding $N+1$ solutions of Eq.~(\ref{poly})
for $z$. Then, for each of them we need to solve a $n_1$-order equation for the $s$, eventually obtaining $n_1$ poles with the same real part. Therefore, the spectrum  displays a further symmetry, splitting in $n_1$ profiles (i.e., the Brillouin zone) each containing $N+1$ poles.

\begin{figure}
\includegraphics[width=1.\linewidth]{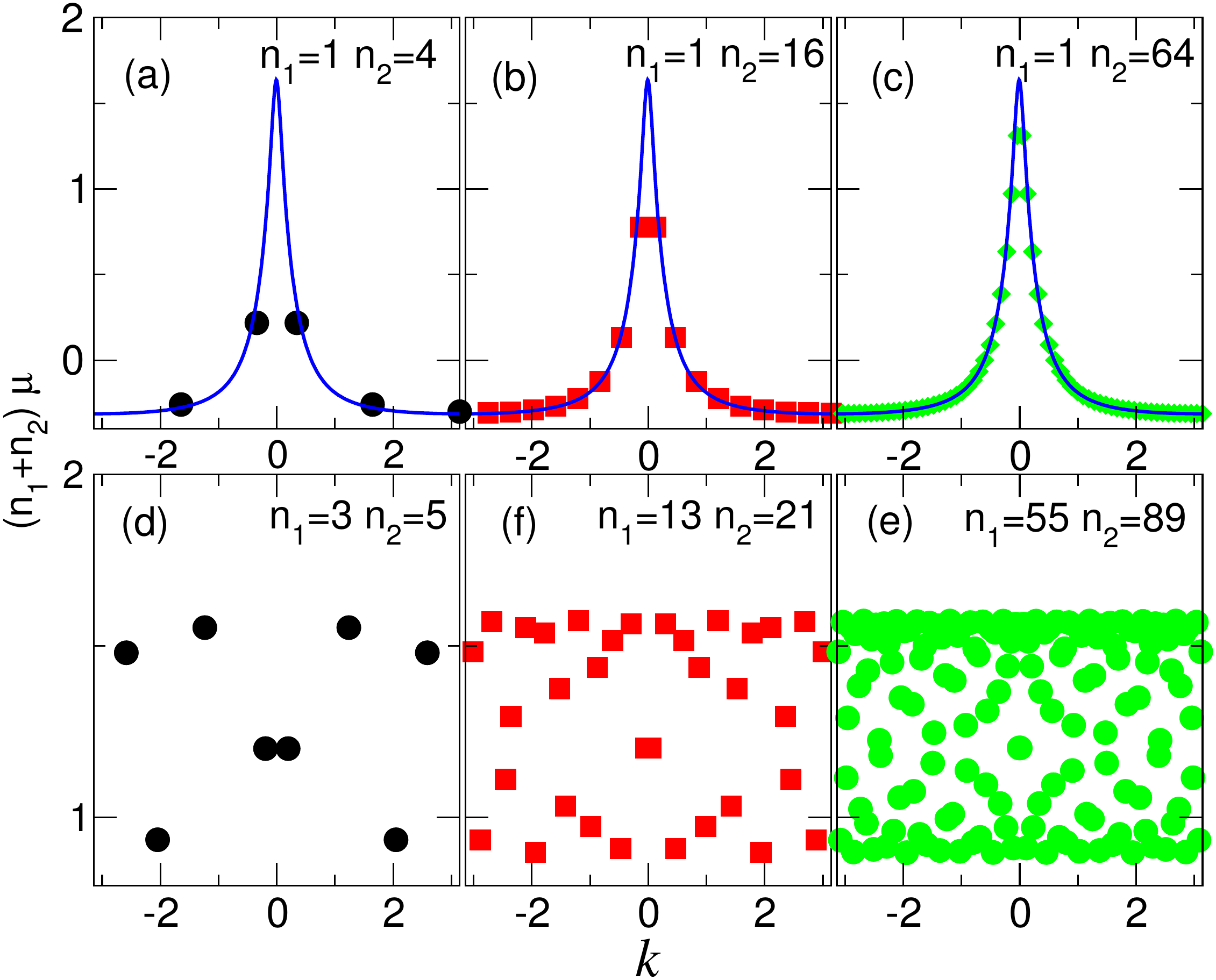}
\caption{The poles in the complex plane for the 1PARA configuration with 
lengths having rational rations, see Eq.(\ref{rational2}) and text. Only the data in the range $|k|<\pi/L_0$ are reported (corresponding to first Brillouin zone for (a)-(c)); the gains are $g_1=1.3,~g_2=0.7$. (a-c) Case $n_2 = N n_1$, solid blue lines is the analytical curve Eq.~(\ref{an1para}); 
(d-e) $n_1$ and $n_2$ being successive value of the 
Fibonacci series. For comparison, the real parts have been multiplied by the 
total length $(n_1+n_2)$.}
\label{1para_rational06}
\end{figure}

As seen in the Fig.\ref{1para_rational06}(a-c), for increasing $N$ the poles distribute along a limit profile after a suitable rescaling of the real part. This can be demonstrated analytically looking for solutions of the type
\begin{equation}
z = e^{-(\mu +ik)L_0} = e^{-(y(x)/N +ix)L_0}~,
\end{equation}
where $x,y$ are independent of $N$ for $N \gg 1$. Substituting in the polynomial equation, we obtain
\begin{equation}
y(x) = \log \Big|(g_1 e^{-ix}-\frac{1}{\sqrt{2}})g_2 \Big| 
-\log\Big|1-\frac{1}{\sqrt{2}} g_1 e^{-ix}\Big|~,
\label{an1para}
\end{equation}

which is depicted as a solid line in the figure showing a very good agreement with the numerical solutions.

Another interesting example is the case in which the integers $n_1$ and $n_2$ 
are pairs of consecutive terms in the Fibonacci series so that the ratio of the 
two physical lengths $L_2/L_1$ approaches the golden ratio $(\sqrt{5}+1)/2$ for increasingly large
$n_1$ and $n_2$. Some of the first steps of the construction are given in Fig.\ref{1para_rational06}(d-e)
that allows to appreciate the increasing complexity of the spectral structure.
It should be noticed that the poles are confined in an horizontal strip in the complex
plane meaning that the growth and decay rates are bounded. 

As an additional remark, note the qualitative similarity of the spectra obtained by 
rational approximation with the numerical solution reported in the left part of 
Fig.\ref{fig:soglia05}, originating from the choice $L_2\approx 2L_1$.  
This confirms that the procedure illustrated in this section
can be useful to understand the spectral structure.

To conclude this Section, we remark that  
the lattice approach can be an effective description of real setups. Indeed, when using arbitrary length fibers and splitters the LANER can be approximated by a sequence of lattices with smaller and smaller periodicity $L_0$.; e.g. $L_0 \to 10^{-r} L_0$ leading to $n_j \to 10^{r} n_j$. As a consequence, the links length is described by an increasing number of digits and the spectrum remains almost the same upon rescaling of the frequency axis (by a factor $\approx 10^{r}$). This approach can be thus considered as an increasingly accurate rational approximation of the spectrum. In a physical system, the ultimate limit for $L_0$ would be represented by the field wavelength; in practice, many effects might limit the observable FSR.

\section{Experiment}

\subsection{Experimental setup}
\label{sec:setup}

\begin{figure*}
\includegraphics[width=0.8\linewidth]{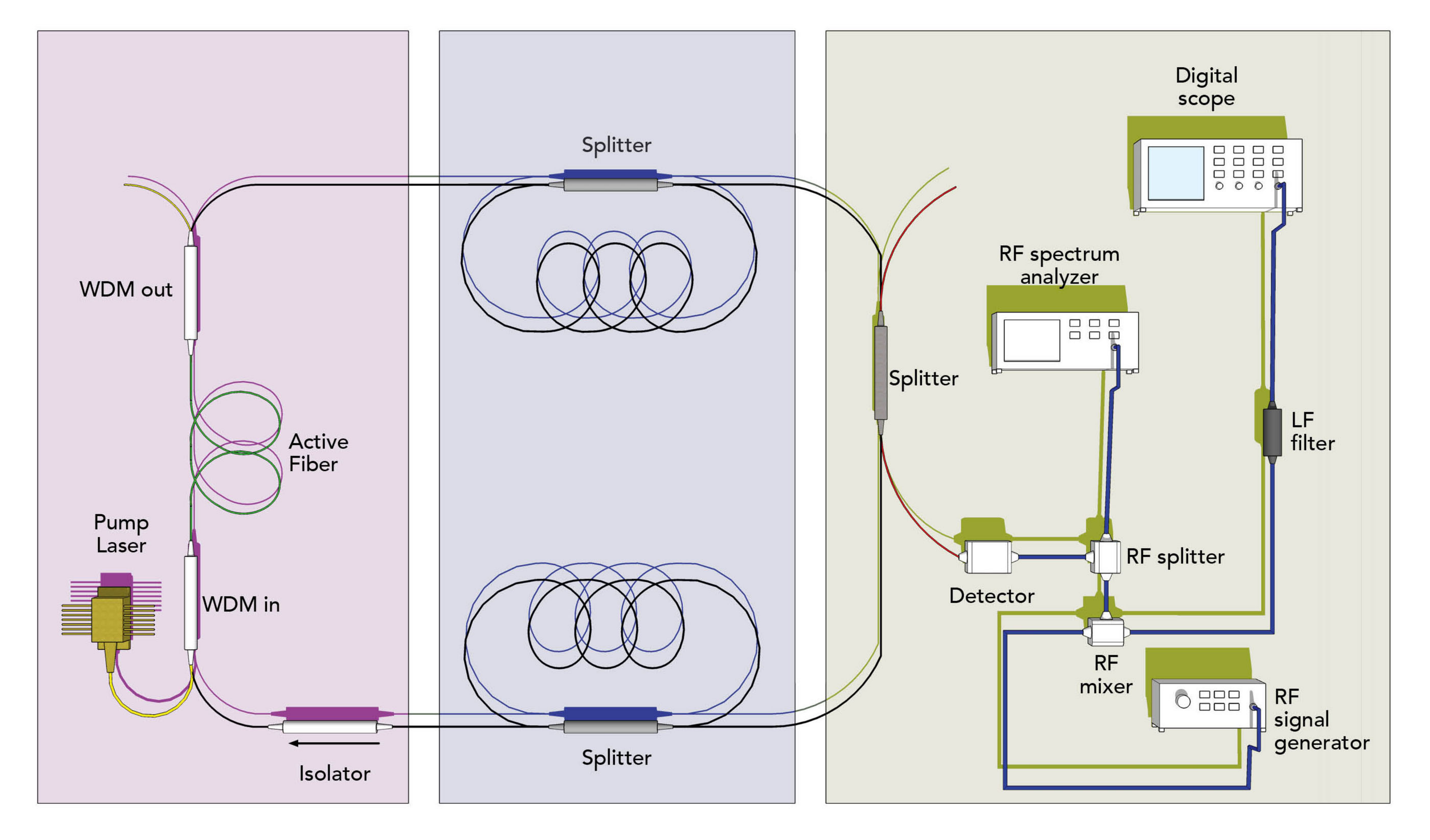}
\caption{Block setup for the case of Fig.\ref{configs04}b (2PARA) of the main text. Left, (oriented) active section composed by a $980~nm$ pumped, $Er^{3+}$ fiber, $WDM$s combining and separating pump and $1.55~\mu m$ signals and an optical isolator. This scheme is represented  as arrowed (red) segments in Fig.1 of the main text.  Center, the rest of the network. Right, the frequency and time-domain detection/acquisition section. }
\label{setup07}
\end{figure*}

In this section, we present the main phenomenology found in the configurations outlined in Fig.\ref{configs04}, focusing on the laser emission, the basic features of the spectrum and leaving more detailed and specific investigations to future studies. Indeed, besides the features of a laser process that can be evidenced already in the DC total intensity a complicated structure of the emission spectrum of a LANER is expected as e.g. a gain is varied. 

As discussed in Section \ref{sec:laner}, a LANER can be physically realized in many ways depending on the components chosen. For instance, a different splitting ratio of the $2 \times 2$ unitary splitters considered before can strongly influence the results obtained in the same topology already in the simplest configurations. 

In the present work, we focus on experimental setups characterized by the following choices and outlined in the block structure of Fig.\ref{setup07}. 

The network is build using single-mode optical fibers and standard, ${\it 2\times2}$ loss-less, reciprocal and matched optical splitters. A ${50:50}$ splitting ratio has been mainly used in the configurations here considered. 

The gain in the active links is provided by $\approx 1.5$~m Erbium-doped optical fiber(s), pumped by $980~nm$ power laser(s). In gain section(s), input and output wavelength de-multiplexers (WDMs) are used to couple/decouple the pump beam to/from the $1550~nm$ signal beam. The unused output port of the WDM-out can be used to monitor the amount of pump power not absorbed by the network. The active links are {\it directed}, i.e. optical isolators assure the unidirectional propagation of the signal beam. This represents a relevant simplification in the expected experimental effects, since possible spatial effects in the active medium (such as spatial hole burning due to the interference between the two counter-propagating fields) are avoided.  The currents of the pump lasers represent the main control parameters of our setup, as they are directly related to the coherent amplification in the active links: an arbitrary number of them can be active simultaneously and independently. A single gain, directed setup for an active link is depicted in the left part of Fig.\ref{setup07}.

The detection is performed by inserting in a network link a 
${\it 2\times2}$ optical coupler with ratio of $50:50$ (symmetric) or $90:10$ (low insertion losses). The split field intensity is collected in the proper direction either with a high bandwidth ($8~GHz$) or low bandwidth ($150~MHz$), high sensitivity PIN photodetectors. The detector current is then sent to a radio-frequency spectrum analyzer and/or a fast digital oscilloscope. A detection/measuring section is illustrated in the right part of Fig.\ref{setup07}.

\subsection{Experimental results}
\label{sec:exp}

\begin{figure}
\includegraphics[width=0.9\linewidth]{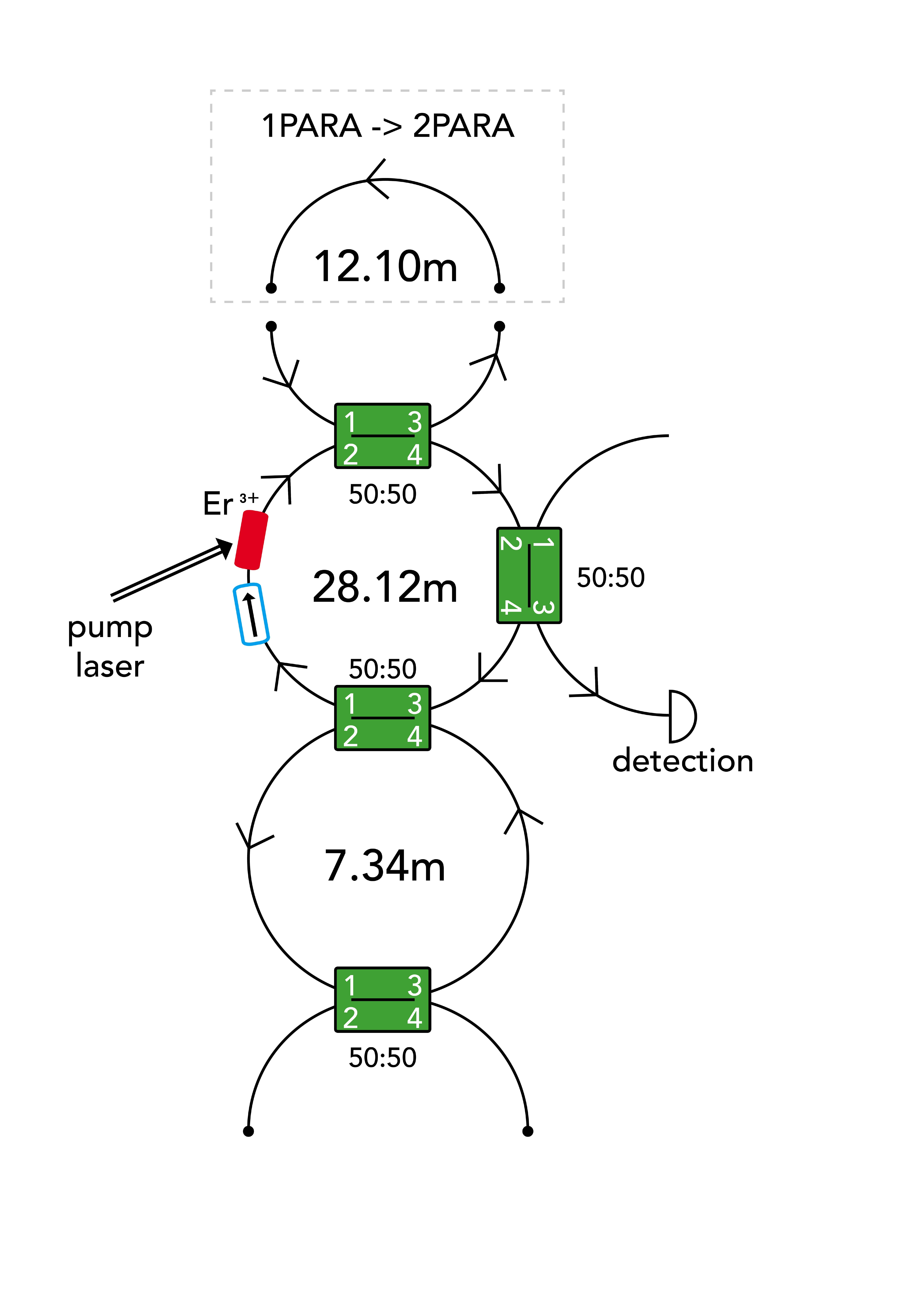}
\caption{Experimental setup for the 1PARA and 2PARA configurations. The un-connected ports of the splitters allow to expand the network without modifications to the existing parts (see the top ring), at the expenses of an higher amount of losses in the simpler arrangements.}
\label{1-2para_exp08}
\end{figure}

The first experimental configurations investigated are depicted in Fig.\ref{1-2para_exp08}. A main ring is built including the active fiber, an optical isolator and three $50:50$ optical splitters; one of them is devoted to bring out the field for the detection. The total length of the ring is measured pumping the active fiber in order to have a laser emission. The measured intensity spectrum, characterized by equally spaced peaks is typical of a ring laser (see e.g. \cite{milonni2010laser}) and allows to determine the cavity length by measuring the peak frequency separation. With the same method, by properly replacing the fiber links it is possible to estimate the lengths of the other rings. In the configuration depicted in the figure, some additional optical splitters have been inserted (such as those in the upper and lower rings) to gain flexibility in the re-organization of the networks links at the expenses of an higher level of losses in the cavity. In particular, we can switch from the 1PARA to the 2PARA configuration by inserting the upper fiber only, while maintaining all the features (e.g. lengths...) of the rest of the network. Notably, in this case the level of losses will decrease because the upper splitter will no more drain power from the network: we therefore expect at least a different value for the threshold current in the two configurations.  

\begin{figure}
\includegraphics[width=0.9\linewidth]{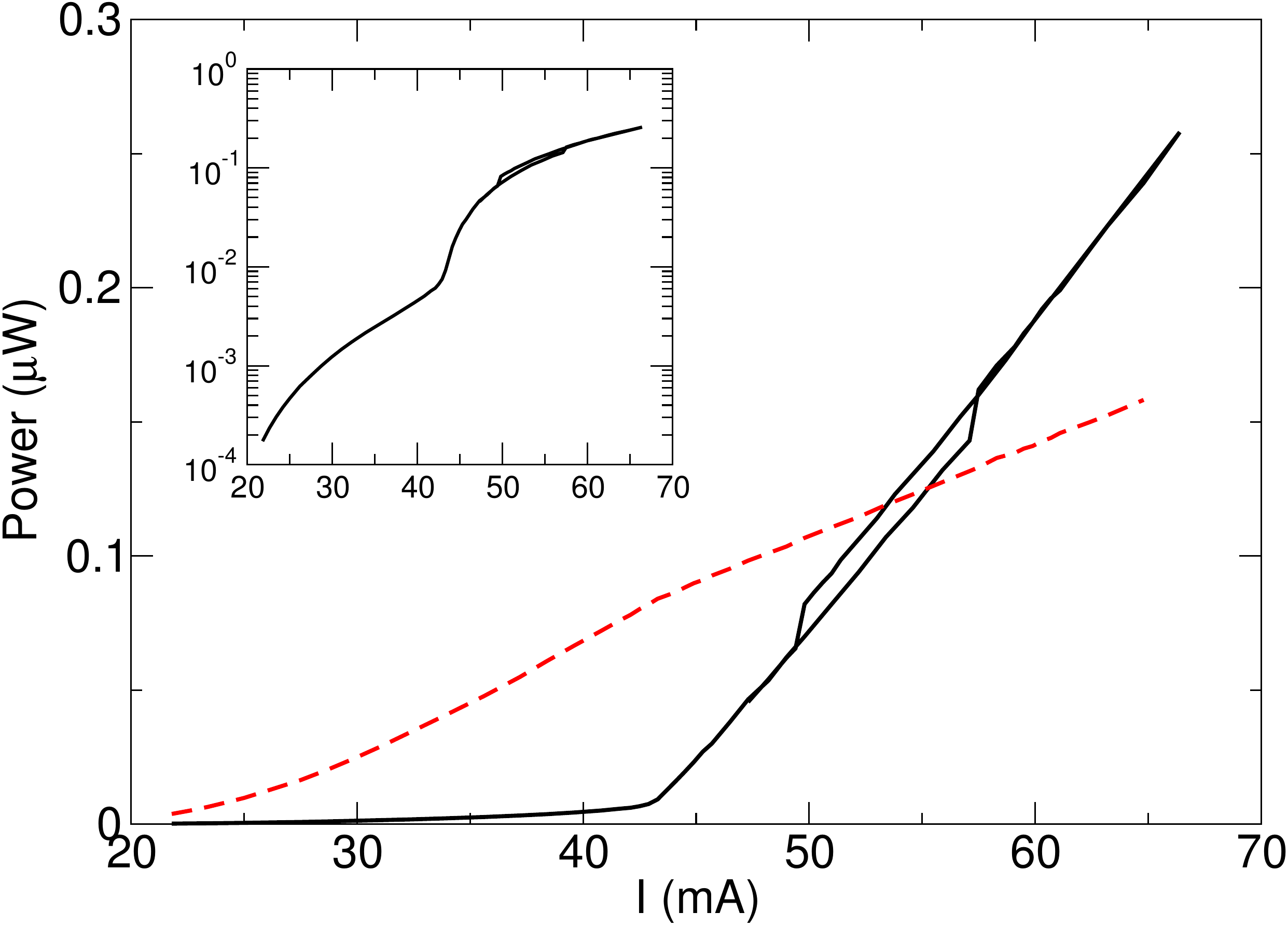}
\caption{Measurement of the LANER emitter power as a function of the laser pump current (black) for the 1PARA configuration. An hysteresis cycle is visible for intermediate pumping power. Inset: the same curve in log scale showing clearly the lasing transition. Red curve: laser pump power measured at the unused port of the WDM-OUT (see Fig.\ref{setup07}); the curve has been rescaled for comparison. The slope change indicates the LANER threshold at $I=43.1~mA$ as well.}
\label{LI09}
\end{figure}

As a first result, we present in Fig.\ref{LI09} the total emission curve as a function of the pump laser current for the 1PARA configuration. This is obtained by a direct measurement of the emitted intensity with a $1.55~\mu m$ optical power meter (black curves), and simultaneously monitoring the fraction of the $980~nm$ pump laser intensity (red curve) exiting from the unused output port of the WDM-OUT (see also Fig.\ref{setup07}). A clear evidence of a laser emission is obtained, with the typical intensity-vs-pump behaviour displaying a threshold point at $I=43.1~mA$. 

\begin{figure}
\includegraphics[width=0.9\linewidth]{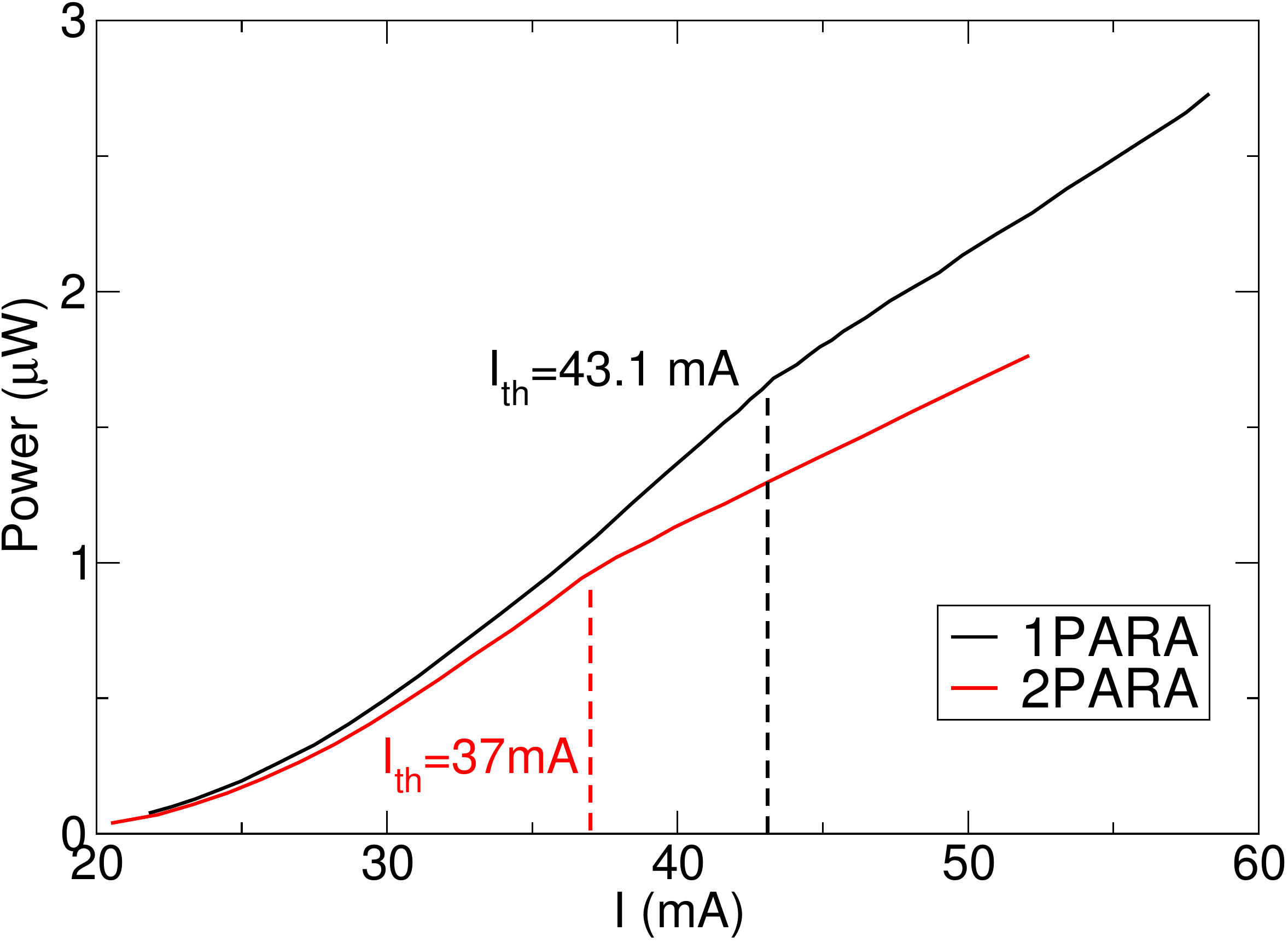}
\caption{Measurement of the laser pump power measured at the unused port of the WDM-OUT (see Fig.\ref{setup07}) for the 1PARA and 2PARA configurations. The corner points indicating the thresholds are marked.}
\label{PL10}
\end{figure}

As seen in the figure, the change in the slope of the red curve indicates a strong change in the energy absorbed by the Erbium active medium as the lasing process starts.  
This method allows an easier estimation of the threshold since the monitoring does not need to reconfigure the detection setup. As an example, we report in Fig.\ref{PL10} the threshold estimation in the 1PARA and 2PARA setups performed in this way.

We also observe that a different level of emission, resulting in an hysteresis cycle when increasing and decreasing the pump laser current is found at intermediate level of pumping (see Fig.\ref{LI09}). Such a phenomenon, typical of complex or multimode laser configurations is expected when nonlinear effects related to far-from threshold pumpings enter into play and will be the object of further investigations.

In the case of class-B lasers \cite{arecchi}, a peculiar feature of the laser dynamics is the interplay between the cavity photon $\tau_p = 1/\kappa$ and the population inversion $\tau_{Er} = 1/\gamma$ lifetimes, producing the so-called {\it relaxation oscillations} (RO) (see e.g. \cite{haken}). They can be observed as noise-induced, damped oscillations in the emitted intensity spectrum and their frequency can be predicted with a linear analysis of the laser rate equations, yielding
\begin{equation}
f_{RO} = {1 \over {2 \pi}}\sqrt{\kappa\gamma\big(I-I_{th}\big)/I_{th} }~.
\label{ro}
\end{equation}

\begin{figure}
\includegraphics[width=0.9\linewidth]{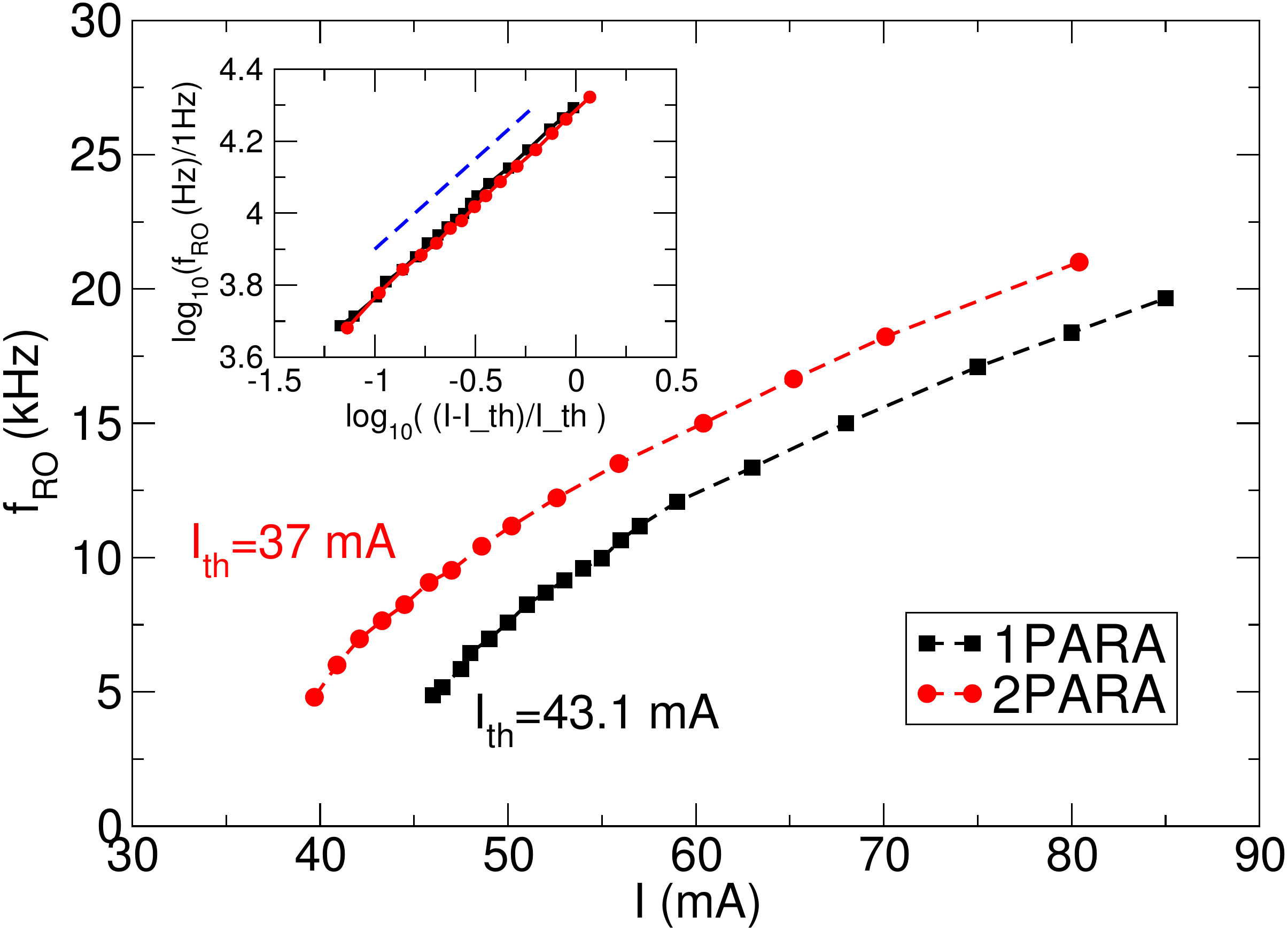}
\caption{Measurement of the relaxation oscillations frequency for the 1PARA and 2PARA configurations as a function of the pump laser current. Inset: rescaled curves for the two configurations (see text). The dashed line is a power law with exponent $1/2$.}
\label{1-2para_RO11}
\end{figure}

We report in Fig.~\ref{1-2para_RO11} the direct measurement of the frequency of RO both in the 1PARA and  2PARA configurations. The curves follow the predicted scaling with the pump, thus confirming the laser nature of the LANER emission. It is interesting to note (as seen in the following) that a strong separation is observed between the dynamical oscillation frequencies, related to the field propagation delays in the network and therefore on the MHz scales, and the {\it natural} frequency related to the interplay of the field and the carriers acting on the KHz scales.

A further information about the system can be obtained by fitting the RO scaling curves and using  \cite{}
\begin{equation}
\tau_p \approx \eta_N T_N~,
\end{equation}

where $T_N = L/v$ is the time corresponding to the total length of the network ($v \approx 200~m\times MHz$ is the light velocity in the fibers) and $\eta_N<1$ is related to the amount of losses in the whole network, obtaining an estimation of the $\gamma$ in the hundredths of Hz in agreement with the known relaxation process time-scales in Er-doped pumped materials \cite{bellemare}.

Most of the dynamical features of the LANER emission can be studied in the spectrum of the emitted intensity. Indeed, in the temporal domain even an almost quasi-periodic behavior easily revealed in the spectrum can be quite difficult to be recognized. Moreover, the power spectrum can be compared with the findings of the linear theory; it is worth remembering, however, that the detected signal spectrum shows the {\it beatings} of the optical modes involved which are estimated by the theory.    

\begin{figure}
\includegraphics[width=1\linewidth]{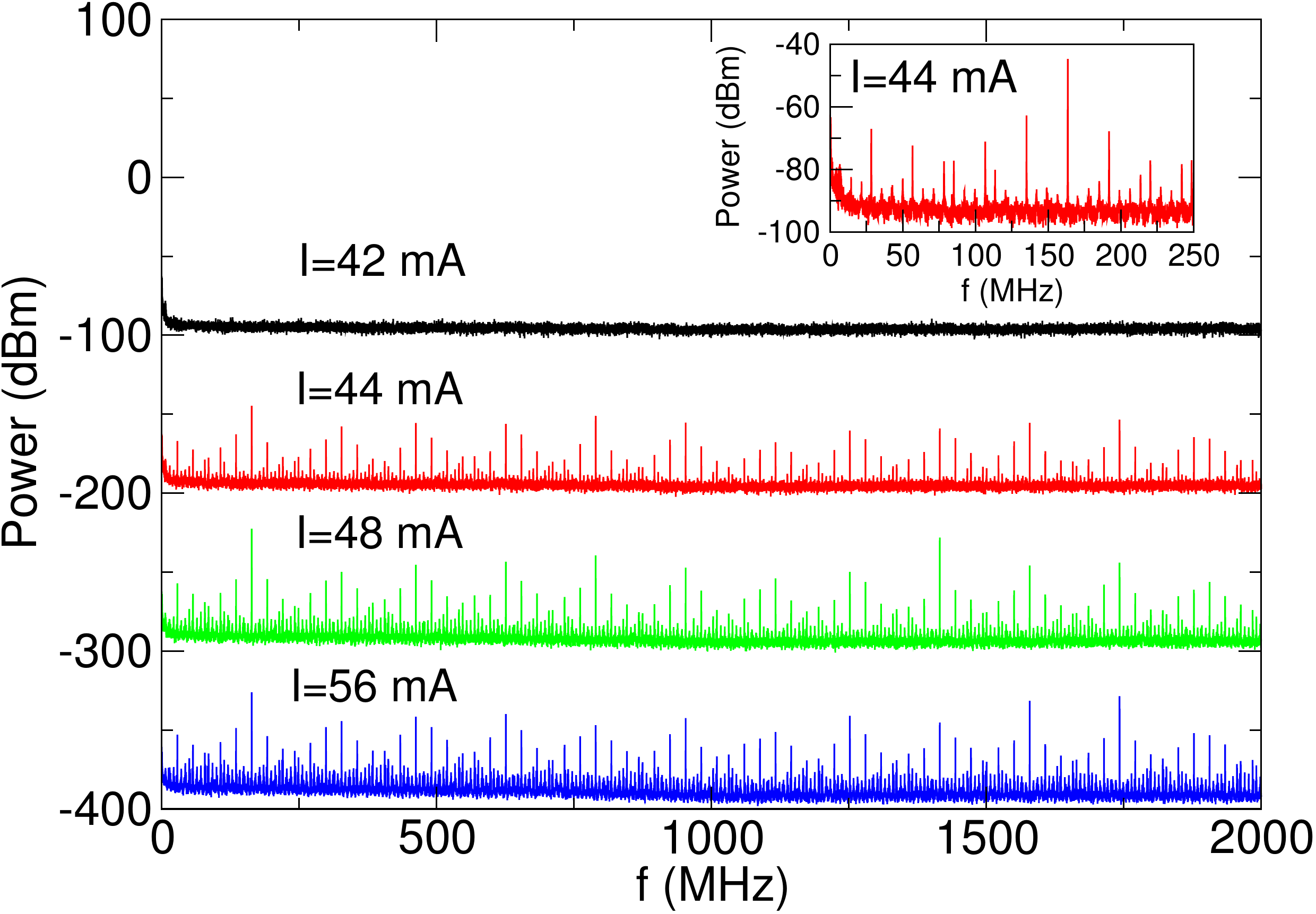}
\caption{Power spectra of the LANER emitted intensity in the 1PARA configuration for increasing values of the pump laser current. The threshold is $I_{th}=43.1~mA$. Each spectrum is shifted vertically by $100~dB$ from the previous one. The inset is a zoom of the spectrum at $I=44~mA$.}
\label{1para_spt12}
\end{figure}

In Fig.~\ref{1para_spt12}, we show the power spectrum of the emitted intensity in the 1PARA configuration for increasing values of the pump laser current. Below the threshold value (at about $I=43.1$~mA as said), no emission is obtained (see the first top spectrum); increasing the pump, the emission starts and many peaks appear, displaying an higher complexity as new or previously unresolved peaks become visible.

As shown in the figure (see also the inset), increasing the pump apparently does not produce relevant changes in the beating spectrum, apart a general increase of the peak height. In particular, the smaller peaks could be generated by nonlinear effects leading to higher harmonics and their corresponding beatings (inter-modulations).

\begin{figure}
\includegraphics[width=1\linewidth]{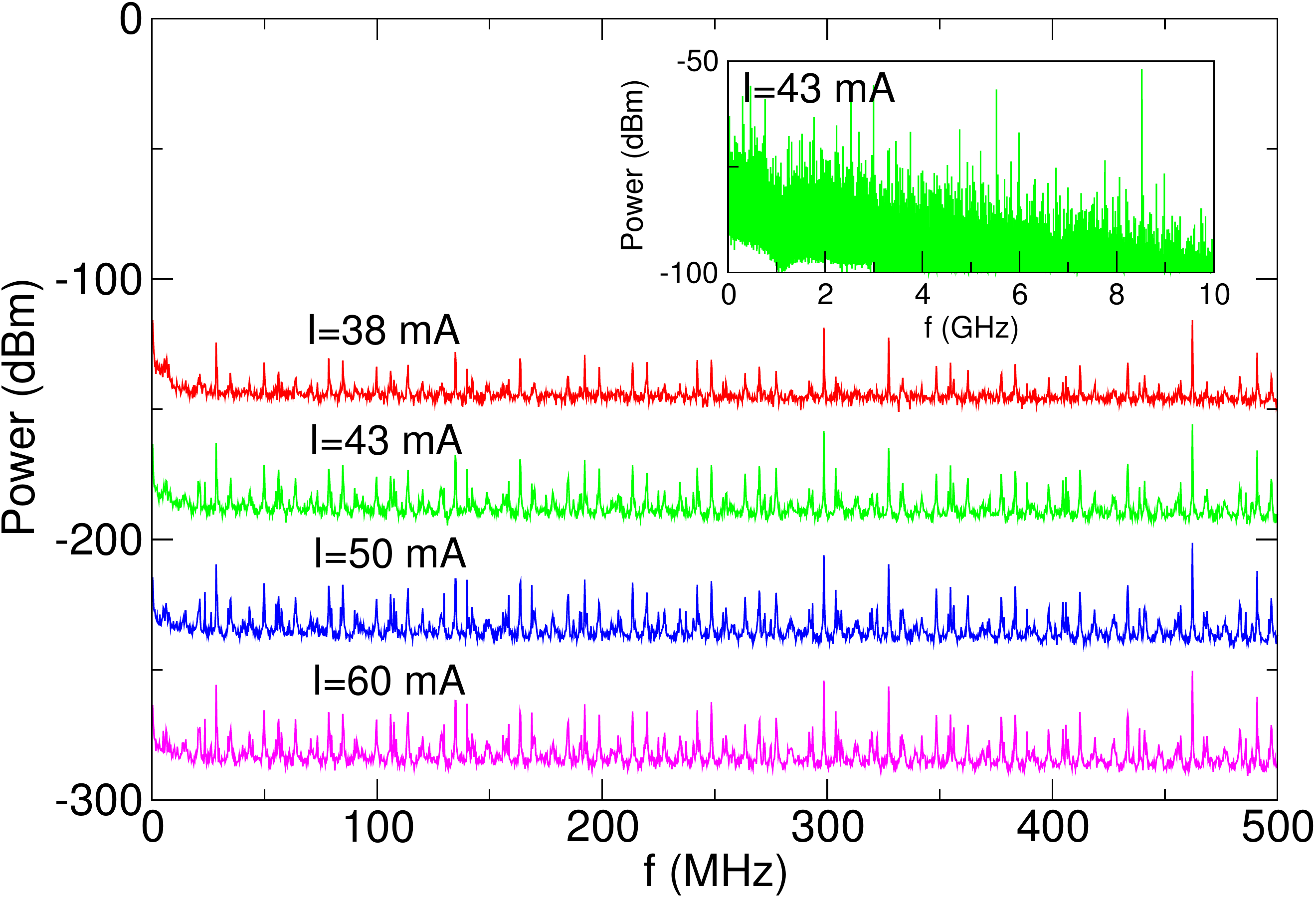}
\caption{Power spectra of the LANER emitted intensity in the 2PARA configuration for increasing values of the pump laser current. The threshold is $I_{th}=37~mA$. Each spectrum is shifted vertically by $50~dB$ from the previous one. The inset is a larger bandwidth spectrum at $I=43~mA$. }
\label{2para_spt13}
\end{figure}

In Fig.\ref{2para_spt13}, we report the same type of measurement for the 2PARA configuration realized by adding the top fiber link as shown in Fig.\ref{1-2para_exp08}. The features of quasi-periodicity of the spectrum is quickly lost increasing the complexity of the topology (see also the inset), at least in the available measurement bandwidth window.

\begin{figure}
\includegraphics[width=0.9\linewidth]{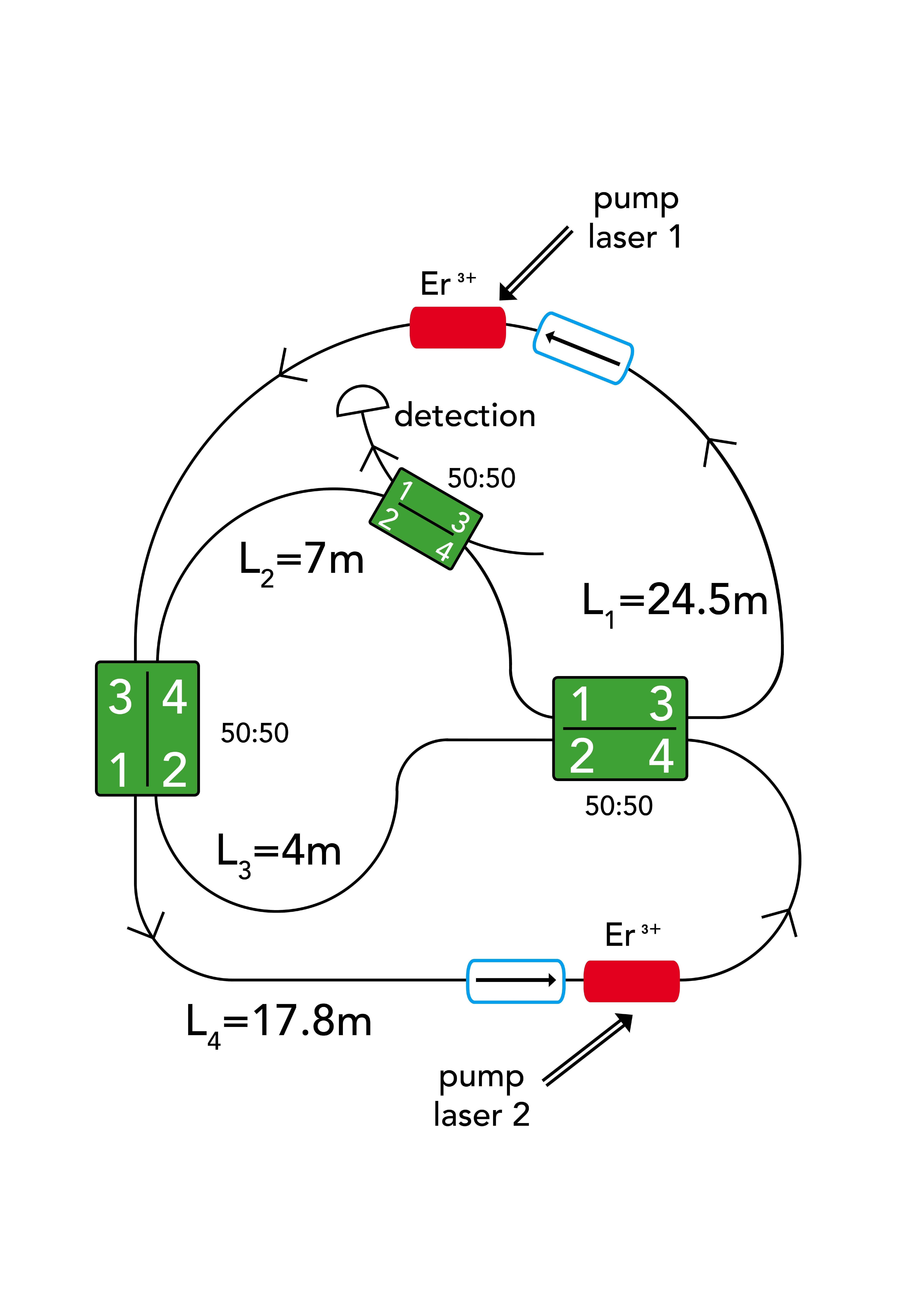}
\caption{Experimental setup for the 2PERP configuration.}
\label{2perp_exp14}
\end{figure}

As a final example, we consider the 2PERP configuration sketched in Fig.\ref{2perp_exp14}. A peculiar feature of this setup is the presence of two active links, each realized as in the left block of Fig.\ref{setup07}. Notably, while in the previous 1PARA and 2PARA setups the optical isolator determined the unidirectional propagation in all the network, this does not occur in the 2PERP even using two of them. In particular, as can be seen in Fig.\ref{2perp_exp14}, the (forced) unidirectional propagation in the links 1 and 4 does not {\it a-priori} leads to same behavior in the links 2 and 3; therefore, both sides of the detection splitter can be used to simultaneously measure the fields in the two directions. 

\begin{figure}
\includegraphics[width=1\linewidth]{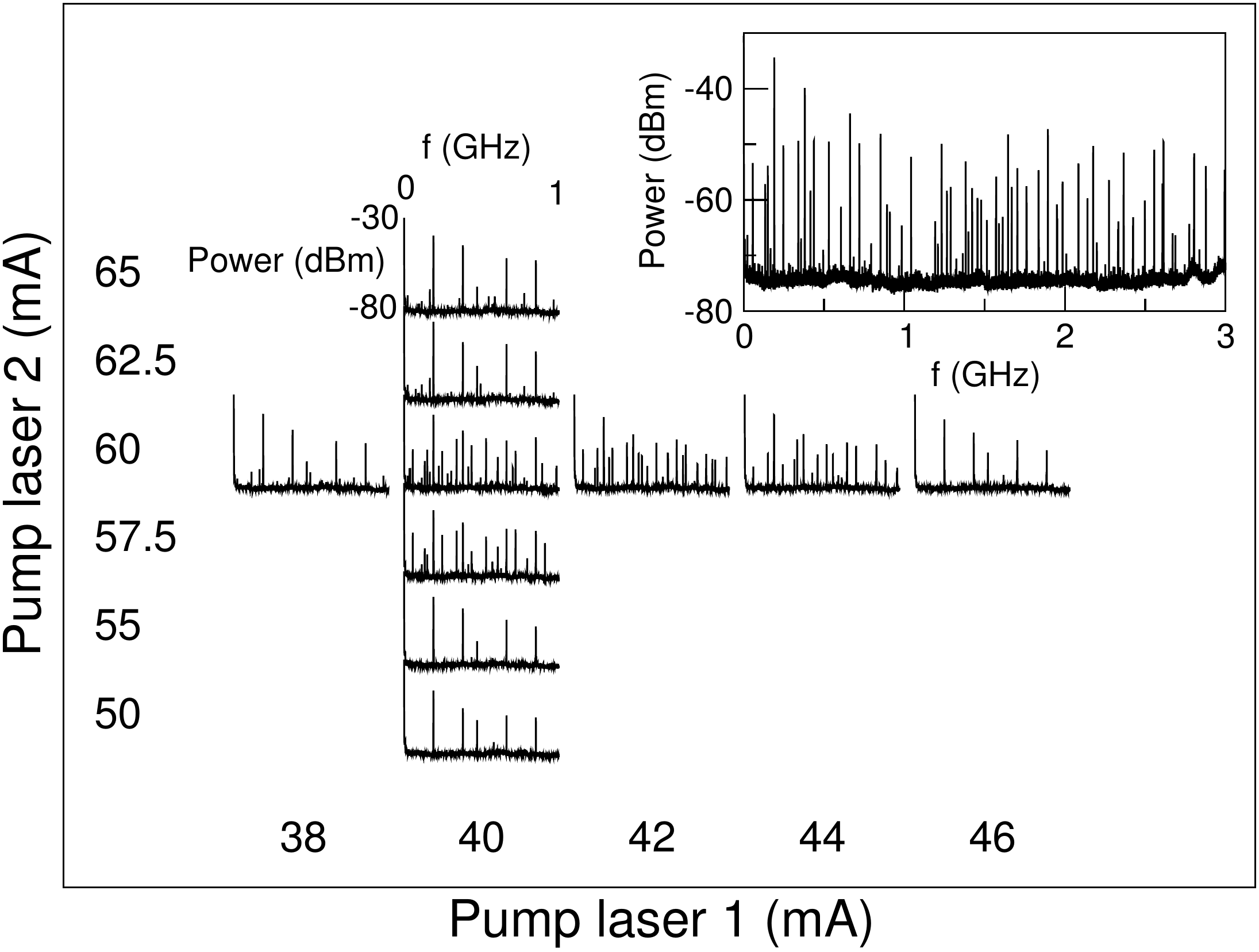}
\caption{Power spectra for the 2PERP configuration when scanning the two pump laser currents. The inset shows a larger bandwidth spectrum for $I_1=40~mA, I_2=60~mA$.}
\label{2perp_scan15}
\end{figure}

In the present work, we limit ourselves to present the effect of the two combined gains as shown in Fig.\ref{2perp_scan15}, in a regime where both the pumped fibers are acting as coherent amplifiers (i.e., the corresponding pump laser currents are large enough). As shown in the figure, a complicated situation occurs, with structured spectra with many peaks found in the two-dimensional parameter space of the gains. Multi-stabilities in the emission are also possible, as a consequence of the complex interplay of the involved fields.      

All the reported phenomenology measured in the three configurations can be compared with those evaluated by computing the linear spectra in Fig.\ref{fig:soglia05}, showing a substantial qualitative agreement. 

\section{Conclusions}

The LANER represents a novel optical scheme generalizing the standard laser setups, both in the geometry of the cavity and in the number of active media providing coherent amplification. In this work we have described the main ingredients of the system, remarking the generality of the approach potentially capable to employ a large variety of components to obtain increasingly complicated configurations. The LANER can also represent as a powerful and flexible experimental framework for the investigation of the dynamics on a network or a graph, where all the control parameters and the variables (related to the involved optical fields) can be easily measured, 

In a linear approach, we have introduced a description based on a general propagation matrix containing the connections (topology) and the lengths and gains (metrics) of the network. A global scattering matrix, describing the transfer properties of the optical couplers connecting the links provides the suitable boundary conditions for the fields. 
The problem defined by Eq.(\ref{char}) for the network matrix, allows the determination of the poles for the wave solutions of the system representing the generalization of the corresponding problem for the standard laser. 
The structure of the solutions is generally quite complex and resembles that found for highly multi-mode laser systems; however, a LANER could be much more complicated because of the freedom in the geometry of the connections.

We have discussed some specific limit cases, of particular interest as they relate to well-known frameworks: the hamiltonian (loss-less cavity) and lattice (all links multiples of the same length) setups, showing the potentiality of the LANER to experimentally realize such difficult schemes. Furthermore, the close connection with the graph theory is suggested with a general method to map a LANER to a directed graph. In this way, an even closer connection with the quantum chaos topics and the problem of amplification and diffusion on a graph has been settled.

To provide examples of the approach, we have studied some simple configurations that can be realized with few couplers, showing the corresponding matrices and numerical solutions of the characteristic problem with the determination of the poles. 

From an experimental point of view, a general discussion of the possible implementations has been provided suggesting that a completely new avenue of studies can be initiated.
We have presented the specific phenomenology of the simpler configurations studied in the theory, giving evidence of the onset of the laser emission above a critical pumping of the active media. 
We reported measurements of the spectra of the emitted intensity where the beatings of the involved modes depict a complicated scenario. 

To conclude, we have presented the LANER as a novel optical scheme, with laser action and characterized by a fully controllable topological disorder induced by the link connections. Since multiple and independent gain sections can be used as well, the LANER would allow rather new investigations, ranging from dynamics on networks of increasing complexity to the effect of the cavity topology on the laser emission. 
We finally remark that the identifications of the key points leading to general results will represent a major challenge of this research. 

\begin{acknowledgments}
We thank F. Cherubini for help in the drawings.
\end{acknowledgments}

\bibliography{bibgraph-long}

\end{document}